\documentclass[twocolumn,10pt,prb,aps,showpacs,superscriptaddress,preprintnumbers,amsmath,amssymb]{revtex4-1}

\usepackage{graphicx}
\usepackage{dcolumn}
\usepackage{bm}
\usepackage{eqnarray}
\usepackage{color}
\usepackage{bigstrut}
\usepackage{tabularx}
\usepackage{float}
\usepackage[]{color}
\usepackage{silence}
\newcolumntype{L}[1]{>{\raggedright\arraybackslash}p{#1}}
\newcolumntype{C}[1]{>{\centering\arraybackslash}p{#1}}
\newcolumntype{R}[1]{>{\raggedleft\arraybackslash}p{#1}}

\begin{document}

\title{Evolution of ground state wave function in CeCoIn$_5$ upon Cd or Sn doping}

\author{K.~Chen}
\altaffiliation{present address: Synchrotron SOLEIL, L'Orme des Merisiers, Saint-Aubin, BP~48, 91192 Gif-sur-Yvette C\'edex, France}
\affiliation{Institute of Physics II, University of Cologne, Z{\"u}lpicher Stra{\ss}e 77, 50937 Cologne, Germany}
\author{F.~Strigari}
  \affiliation{Institute of Physics II, University of Cologne, Z{\"u}lpicher Stra{\ss}e 77, 50937 Cologne, Germany}
\author{M.~Sundermann}
  \affiliation{Institute of Physics II, University of Cologne, Z{\"u}lpicher Stra{\ss}e 77, 50937 Cologne, Germany}
 \affiliation  {Max Planck Institute for Chemical Physics of Solids, N{\"o}thnizer Stra{\ss}e 40, 01187 Dresden, Germany}
\author{Z.~Hu}
 \affiliation  {Max Planck Institute for Chemical Physics of Solids, N{\"o}thnizer Stra{\ss}e 40, 01187 Dresden, Germany}
\author{Z.~Fisk} 
  \affiliation{Department of Physics and Astronomy, University of California, Irvine, California 92697, USA} 
\author{E.~D.~Bauer} 
  \affiliation{Los Alamos National Laboratory, New Mexico 87545, USA} 
\author{P.~F.~S.~Rosa} 
 \affiliation{Los Alamos National Laboratory, New Mexico 87545, USA} 
\author{J.~L.~Sarrao} 
  \affiliation{Los Alamos National Laboratory, New Mexico 87545, USA} 
\author{J.~D.~Thompson} 
  \affiliation{Los Alamos National Laboratory, New Mexico 87545, USA} 
\author{J.~Herrero-Martin} 
  \affiliation{ALBA Synchrotron Light Source, E-08290 Cerdanyola del Vall$\grave{e}$s, Barcelona, Spain} 
\author{E.~Pellegrin} 
  \affiliation{ALBA Synchrotron Light Source, E-08290 Cerdanyola del Vall$\grave{e}$s, Barcelona, Spain}  
\author{D.~Betto}
  \affiliation{European Synchrotron Radiation Facility (ESRF), B.P. 220, 38043 Grenoble C\'edex, France}  
\author{K.~Kummer}
  \affiliation{European Synchrotron Radiation Facility (ESRF), B.P. 220, 38043 Grenoble C\'edex, France}  
\author{A.~Tanaka}
  \affiliation{Department of Quantum Matter, Graduate School of Advanced Sciences of Matter, Hiroshima University, Higahsi-Hiroshima 739-85430, Japan} 
\author{S.~Wirth}
 \affiliation  {Max Planck Institute for Chemical Physics of Solids, N{\"o}thnizer Stra{\ss}e 40, 01187 Dresden, Germany}
\author{A.~Severing}
  \affiliation{Institute of Physics II, University of Cologne, Z{\"u}lpicher Stra{\ss}e 77, 50937 Cologne, Germany}
	\affiliation  {Max Planck Institute for Chemical Physics of Solids, N{\"o}thnizer Stra{\ss}e 40, 01187 Dresden, Germany}

\date{\today}

\begin{abstract}
We present linear polarization-dependent soft x-ray absorption spectroscopy data at the Ce $M_{4,5}$ edges of Cd and Sn doped CeCoIn$_5$. The 4$f$ ground state wave functions have been determined for their superconducting, antiferromagnetic and paramagnetic ground states. The absence of changes in the wave functions in CeCo(In$_{1-x}$Cd$_x$)$_5$ suggests the 4$f$\,--\,conduction electron ($cf$) hybridization is not affected by globally Cd doping, thus supporting the interpretation of magnetic droplets nucleating long range magnetic order. This is contrasted by changes in the wave function due to Sn substitution. Increasing Sn in CeCo(In$_{1-y}$Sn$_y$)$_5$ compresses the 4$f$ orbitals into the tetragonal plane of these materials, suggesting enhanced $cf$ hybridization with the in-plane In(1) atoms and a homogeneous altering of the electronic structure. As these experiments show, the 4$f$ wave functions are a very sensitive probe of small changes in the hybridization of 4$f$ and conduction electrons, even conveying information about direction dependencies. 
  
\end{abstract}


\maketitle

\section{Introduction}
The physics of heavy fermion compounds is driven by competition of two effects, the inter-site Ruderman-Kittel-Kasuya-Yosida interaction and the on-site Kondo screening, with the latter resulting from the hybridization of 4$f$ and conduction electrons ($cf$ hybridization).  In a temperature vs exchange interaction ($J_{ex}$) phase diagram, a magnetically ordered state is formed for small $J_{ex}$ and a non-magnetic Kondo singlet ground state for large $J_{ex}$.\cite{Doniach1977} These two ground states are separated by a quantum critical point (QCP) in the vicinity of which often unconventional superconductivity occurs. The phase diagrams of heavy fermion compounds are therefore interesting playgrounds for studying unconventional superconductivity competing with magnetic ground states as well as the phenomenon of quantum criticality.\cite{Hilbert2007, Stockert2011,Wirth2016} 

Here the Ce$M$In$_5$ compounds with $M$\,=\,Co, Rh and Ir are model systems because their phase diagrams comprise unconventional superconductivity (SC) and  antiferromagnetism (AF) as well as the coexistence of these phases (SC+AF).\cite{Thompson2012} CeCoIn$_5$ and CeIrIn$_5$ are ambient pressure superconductors with transition temperatures $T_c$ = 2.3\,K and 0.4\,K, respectively,  whereas CeRhIn$_5$ is an antiferromagnet with $T_N$\,=\,3.8\,K that becomes superconducting with applied pressure. The interplay of superconductivity and magnetism is manifested in the coexistence of phases in CeRh$_{1-\delta}$Ir$_{\delta}$In$_5$ as well as in the pressure induced superconductivity of CeRhIn$_5$. CeRhIn$_5$ exhibits a magnetic quantum critical point (QCP) with applied pressure. Evidence for a QCP is also observed in CeCoIn$_5$ even though AF order is not found under ambient conditions but appears to be close by.\,\cite{Petrovic2001,Hu2012} The normal state of CeCoIn$_5$ exhibits non-Fermi liquid scaling with specific heat divided by temperature C/$T$\,$\propto$\,-ln$T$ and resistivity $\rho$($T$)\,$\propto$\,$T$.

\begin{figure}
	\centering
   \includegraphics[width=0.98\columnwidth]{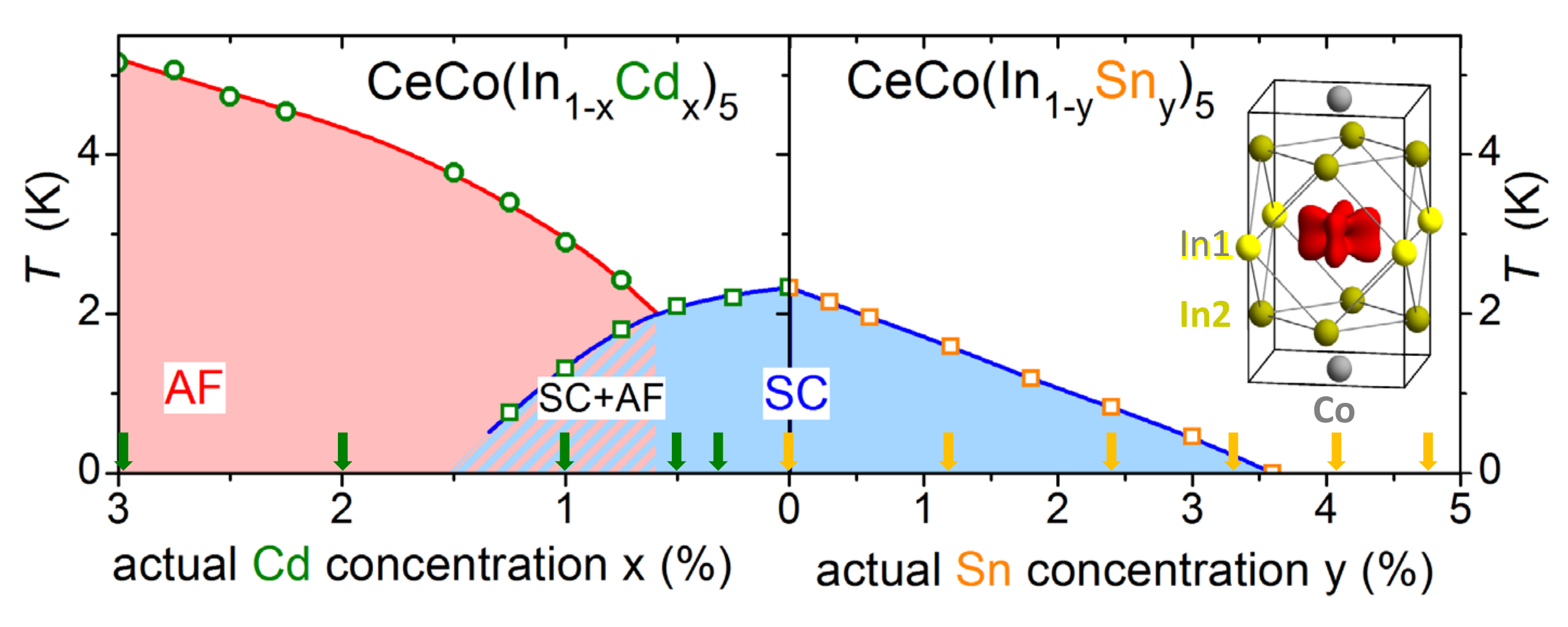}
\caption{(Color online) Temperature\,--\,doping phase diagram of CeCo(In$_{1-x}$Cd$_x$)$_5$ and CeCo(In$_{1-y}$Sn$_y$)$_5$, adapted from.\cite{Bauer2006,Ramos2010} Here, concentrations are given in\,\%. The inset shows the crystallographic structure of CeCoIn$_5$ with Ce (red orbital) in the middle, Co (grey), In(1) (yellow), and In(2) (dark yellow).}
		\label{Fig_1}
\end{figure}

The Ce$M$In$_5$ compounds form in the tetragonal HoCoGa$_5$ structure that can be understood as alternating CeIn$_3$ and $M$In$_2$ layers with two non-equivalent In sites, In(1) in-plane and In(2) out-of-plane with Ce (see inset of Fig.\,\ref{Fig_1}). What drives the ground state properties in these isostructural compounds is an ongoing debate. It had been speculated that lattice anisotropy, i.e. the ratio of the tetragonal $c$ and $a$ lattice constants, may drive the ground state properties because the superconducting transition temperature $T_c$ varies linearly with $c/a$ in the Ce as well as Pu 115 families. Yet, this hypothesis makes no statement about the antiferromagnetic members of the families.\cite{Pagliuso_2002a,Bauer2004}

At a more fundamental level, the $f$-electron ground state orbital should have an impact on ground state properties.\cite{Pagliuso_2002,Shim2007,Burch2007,Haule2010,Pourovskii2014} And indeed, we have shown by soft x-ray absorption (XAS) experiments on CeCoIn$_5$ and the CeRh$_{1-\delta}$Ir$_{\delta}$In$_5$ (0\,$\leq$$\delta$\,$\leq$\,1) substitution series that the 4$f$ ground state wave function correlates with the phase diagram:\cite{Willers2015} the 4$f$ orbitals of superconducting CeCoIn$_5$, CeIrIn$_5$, and Ir rich CeRh$_{1-\delta}$Ir$_{\delta}$In$_5$ are elongated in the $c$ direction with respect to the ground state wave functions of antiferromagnetic members of the CeRh$_{1-\delta}$Ir$_{\delta}$In$_5$ series. The enhanced hybridization of Ce with out-of-plane In(2) ions in the $M$\,=\,Co and the Ir rich samples provides a rational explanation for why they favor superconductivity.\cite{Shim2007,Burch2007,Haule2010} This result was intriguing because it suggests that another parameter, namely the 4$f$ ground state wave function, has to be accounted for when modeling these materials with the goal to eventually predict superconductivity. 

To generalize these findings and further elucidate the role of the 4$f$ orbital on ground state properties, we consider the phase diagrams of CeCo(In$_{1-x}$Cd$_x$)$_5$ and CeCo(In$_{1-y}$Sn$_y$)$_5$ where In is substituted with Cd\,\cite{Pham2006,Nicklas2007,Nair2010,Seo2014} or Sn\,\cite{Bauer2005a,Daniel2005,Bauer2005,Donath2006,Bauer2006,Ramos2010} (see Fig.\,\ref{Fig_1}). Substitutions on the In sites have the  advantage -- in contrast to CeRh$_{1-\delta}$Ir$_{\delta}$In$_5$ -- that the impact on the lattice is minor.

Both phase diagrams have been carefully investigated. Electron doping with Sn and hole doping with Cd have very different consequences on $cf$ hybridization.\cite{Sakai2015} Cd leads to a reduction of $T_c$ (SC phase) and at only 0.6\% Cd antiferromagnetic order appears. There is a region of phase coexistence (SC+AF) for x\,=\,0.6\,--\,1.55\% in which the superconducting transition temperature $T_c$ decreases further and the N$\acute{e}$el temperature $T_N$ increases. Beyond 1.55\% Cd doping superconductivity is suppressed and $T_N$ increases up to 4.6\,K at 3\% Cd (AF phase).\cite{Pham2006,Nicklas2007,Nair2010,Seo2014}  This is contrasted by substitution with Sn. Upon increasing Sn substitution $T_c$ decreases almost linearly and superconductivity disappears completely for y\,$\geq$\,3.5\%. The $cf$ hybridization increases with Sn concentration according to a shift of the maximum in the electrical resistivity towards higher temperatures.\cite{Bauer2005a,Daniel2005,Bauer2005,Donath2006,Bauer2006,Ramos2010} In a nuclear quadrupole investigation\cite{Sakai2015} Sakai \textsl{et al.} showed that Cd substitution creates a heterogeneous electronic state in which unscreened magnetic moments are induced in the local vicinity of Cd dopants, but the majority of electronic states is unaffected. In contrast, Sn substitution enhances $cf$ hybridization uniformly throughout the crystal.
 
At small concentrations of Sn and Cd dopants, there is negligible change in lattice-structure parameters\cite{Booth2009} but a large change in ground state, making the two substitution series well suited for investigating the interplay of ground state wave function and ground state properties. In the following we present results of a soft XAS investigation at the Ce $M_{4,5}$ edge of the Cd and Sn substitution series. Soft XAS with linear polarized light is a powerful tool for investigating the evolution of the ground state wave function since it provides information about changes in the wave functions with great accuracy.\cite{Hansmann2008,Willers2010,Willers2015} We find that Cd doping does not alter the wave function whereas Sn doping does.

\begin{figure*}
	\centering
	 \includegraphics[width=1.8\columnwidth]{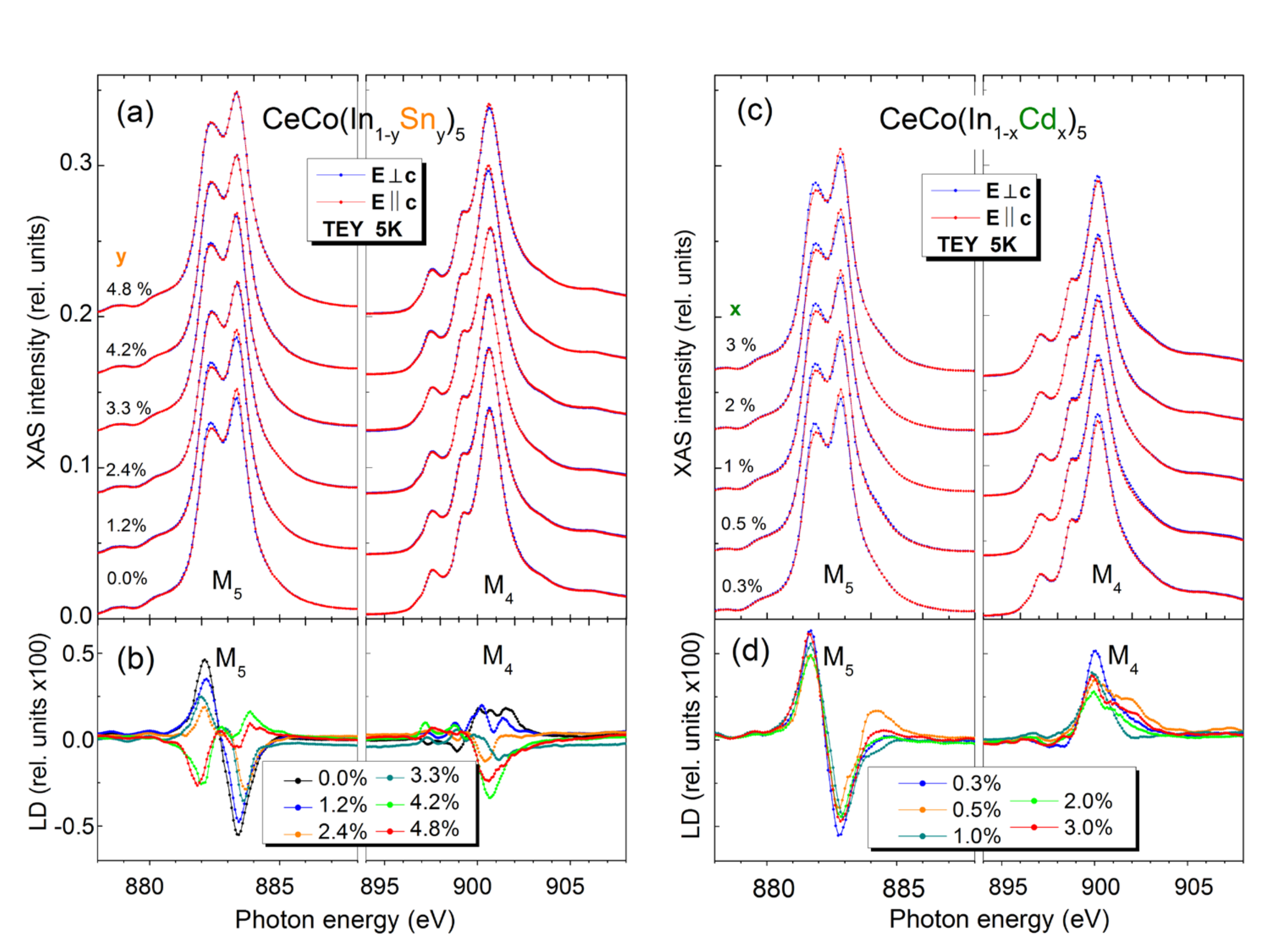}
	\caption{(Color online) (a) and (c): Linear polarized $M_{4,5}$ x-ray absorption data of CeCoIn$_5$ and all concentrations of CeCo(In$_{1-y}$Sn$_y$)$_5$ and CeCo(In$_{1-x}$Cd$_x$)$_5$. For clarity the spectra are shown with a vertical off-set. (b) and (d): Linear dichroism LD\,=\,I($E\| c$)-I($E\bot c$) for all Sn and Cd concentrations, respectively. The LD have been been smoothed.}
		\label{Fig_2}
\end{figure*}

\begin{figure*}
	\centering
  \includegraphics[width=1.99\columnwidth]{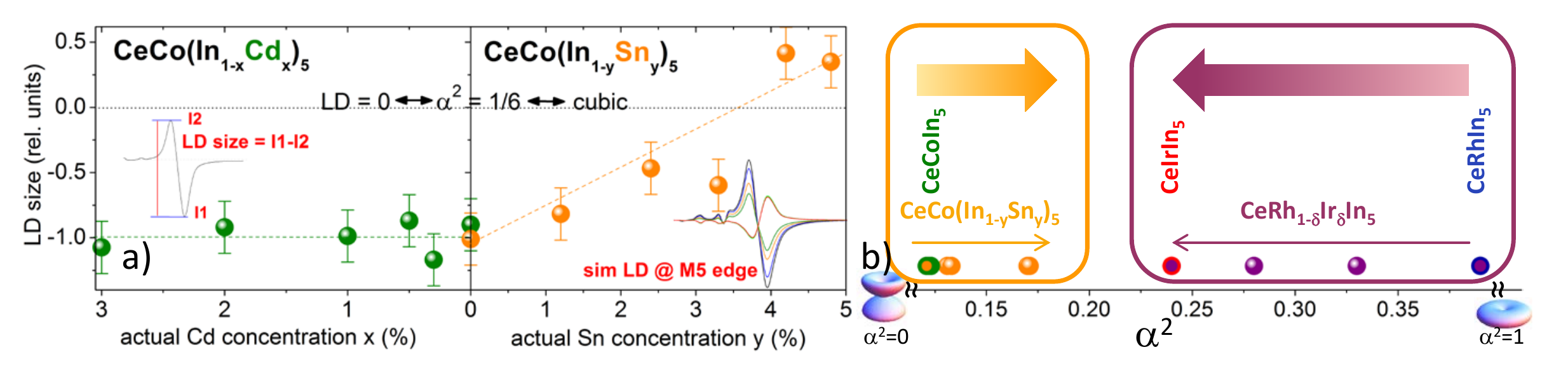}
\caption{(Color online) Evolution of LD-size and $\alpha^2$ as a function of material composition. (a) Size of LD at the M$_5$ edge -- definition in the inset on the left -- versus Cd and Sn concentration. The sign of the LD is defined as in Ref.\,\onlinecite{Willers2015}. The broken lines are guides to the eye. The inset on the right shows the simulated LD-size that follows the orange line. The color code for the concentrations is as in Fig.\,2(b). (b) $\alpha^2$ values of CeCoIn$_5$, CeCo(In$_{1-x}$Cd$_x$)$_5$ (green) and CeCo(In$_{1-y}$Sn$_y$)$_5$ (orange), plus the values of CeRhIn$_5$ (blue), CeIrIn$_5$ (red) and CeRh$_{1-\delta}$Ir$_{\delta}$In$_5$ from our previous study,\cite{Willers2015} (purple). The thick arrows indicate the increase of $cf$-hybridization within the respective substitution series. For $\alpha^2$\,=\,0 and 1 the corresponding orbitals are shown. The orange and red frames indicate that the two substitution studies should be seen independently due to the different direction dependence of hybridization.
}
		\label{Fig_3}
\end{figure*}

\section{Experiment and Analysis}
Single crystals of CeCo(In$_{1-x}$Cd$_x$)$_5$ and CeCo(In$_{1-y}$Sn$_y$)$_5$ were grown out of an In flux as described earlier.\cite{Pham2006,Bauer2008} Microprobe measurements show that the actual Cd and Sn content in crystals is approximately 10\% and 60\% of their nominal concentration in the flux, respectively.\cite{Pham2006,Bauer2005} Throughout, we use the actual concentrations x and y, expressed as a percentage, for Cd and Sn, respectively.  X-ray absorption fine structure (XAFS) studies find that Cd and Sn preferentially occupy In(1) sites, with Sn at In(1) roughly twice as likely as Cd at In(1).\cite{Daniel2005a,Booth2009} Measurements were conducted on samples with Sn substitution of y\,=\,1.2, 2.4, 3.3, 4.2, and 4.8\% and Cd substitution of x\,=\,0.3, 0.5, 1, 2, and 3\% (see orange and green arrows in Fig.\,\ref{Fig_1}). Consequently, ground states in both phase diagrams are well covered.

X-ray linear dichroism experiments at the Ce-M$_{4,5}$-edges (880-904\,eV) were performed on two soft x-ray beamlines, BL29 BOREAS at the ALBA synchrotron in Barcelona, Spain\,\cite{Barla2016} and ID32\,\cite{Kummer2016} at the European Synchrotron Radiation Facility (ESRF) in Grenoble, France. At ALBA the energy resolution was set to 200\,meV and at ESRF to 150\,meV. The spectra were recorded using the total electron yield (TEY) method by measuring the sample drain current in a chamber with a vacuum base pressure of 2$\cdot$10$^{-10}$\,mbar (ALBA) and 10$^{-11}$\,mbar (ESRF). The samples were cleaved \textsl{in situ} to obtain clean sample surfaces. The single-crystalline samples were aligned so that a surface with a (100) normal vector was exposed to the beam. This way both polarizations, $E\| c$ and $E\bot c$, could be measured without having to rotate the sample. The measurements were performed at base temperatures with the sample surface at 5\,$\pm$1\,K.

All spectra are normalized to the integrated intensities so that the difference of the two polarizations (linear dichroism) is directly comparable. Data simulation was performed with the full multiplet code XTLS\,\cite{Tanaka1994} as described previously.\cite{Hansmann2008,Willers2010,Willers2015} We recap that the ground state 4$f$ wave function of Ce in the tetragonal point symmetry of Ce$M$In$_5$ is $\Gamma_7^{\pm}$\,=\,$\alpha$$|\pm5/2\rangle\,\pm\,\sqrt{1-\alpha^2}$$|\mp\,3/2 \rangle$ and that we use $\alpha^2$ for parametrizing the ground state wave function. At 5\,K excited crystal-field levels are not populated\,\cite{Christianson2002,Chrstianson2004,Willers2010} and hence the XAS experiment only probes the ground state.

\section{Results}
Figures\,\ref{Fig_2}\,(a-d) show XAS spectra of the M$_5$ edge together with the linear dichroism, LD\,=\,I($E\| c$)-I($E\bot c$), for all Cd and Sn concentrations. We start by comparing undoped CeCoIn$_5$ (0.0\% Sn) to the sample with the highest Sn concentration (4.8\% Sn) in Fig.\,\ref{Fig_2}\,(a). For CeCoIn$_5$ the linear polarized data agree very well with our previous findings of a fairly small LD of opposite sign found in CeRhIn$_5$ and CeIrIn$_5$.\cite{Willers2010} Comparing the spectra of undoped CeCoIn$_5$ spectra with that of the 4.8\% Sn sample shows a decrease in polarizations. Looking in greater detail at the LD on an expanded scale (compare the black and red dotted lines in Fig.\,\ref{Fig_2}\,(b)), it becomes apparent that the difference in the spectra has changed sign. In absolute terms the effect is very small; however, Fig.\,\ref{Fig_2}\,(b) also shows the difference spectra for all interim Sn concentrations and clearly there is a systematic trend. The LD decreases with increasing y and eventually changes sign for the highest Sn concentrations of 4.2 and 4.8\%.

\begin{figure}
	\centering
  \includegraphics[width=0.99\columnwidth]{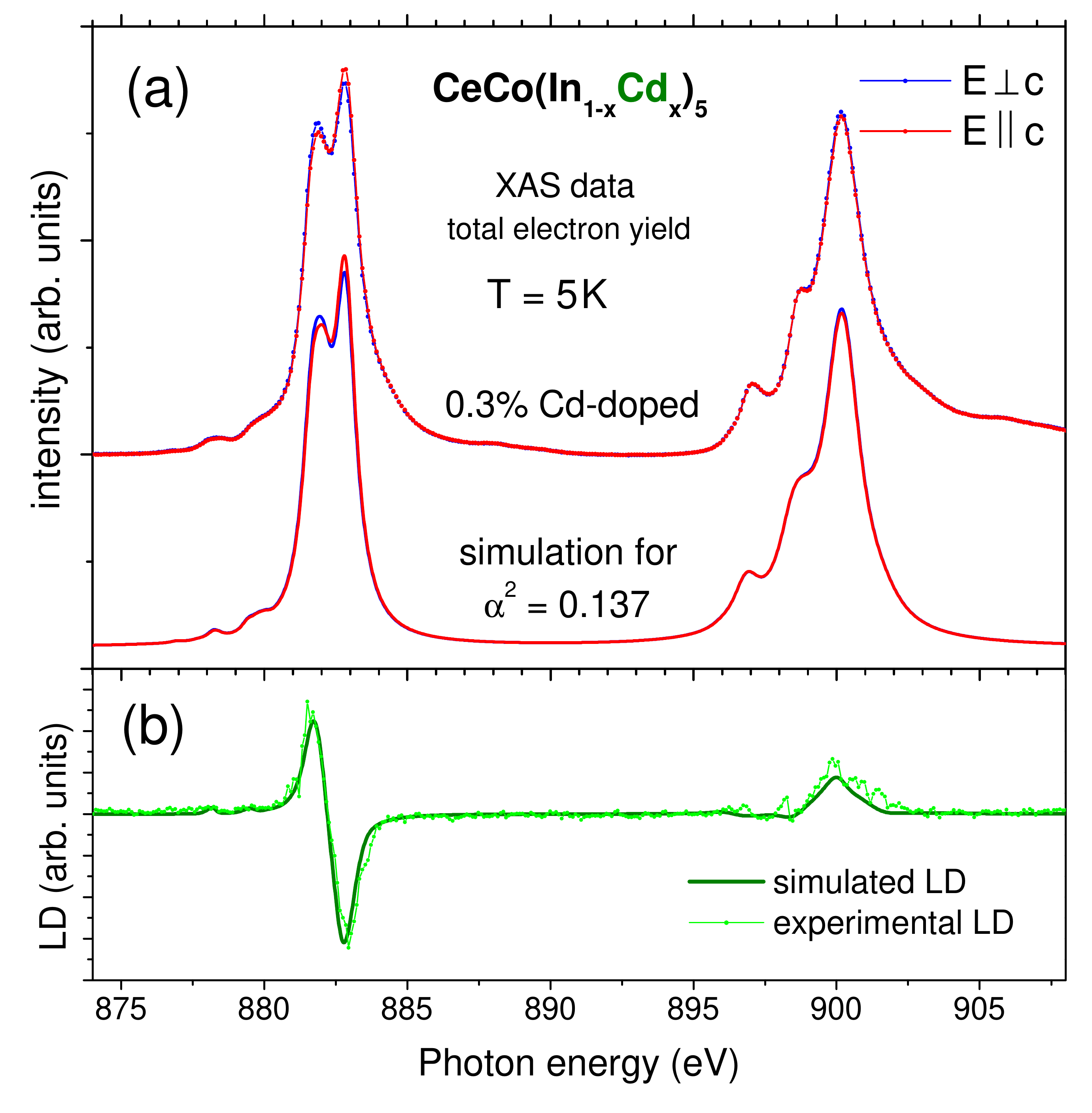}
	\caption{(Color online) (a) $M_{4,5}$ x-ray absorption of CeCo(In$_{1-x}$Cd$_x$)$_5$ for x\,=\,0.3 and the simulation for $\alpha$$^2$\,=\,0.137. (b) Experimental and simulated linear dichroism.}
		\label{Fig_4}
\end{figure}

Figures\,\ref{Fig_2}\,(c) and (d) present the XAS spectra plus the LD on an expanded scale for all Cd substitutions (0.3 to 3\%). There is no apparent change in the polarization dependence. As illustrated in Fig.\,\ref{Fig_4} for 0.3\% Cd, experimental XAS and LD spectra can be simulated well with $\alpha^2$\,=\,0.137 which is in reasonable agreement with $\alpha^2$\,=\,0.13 for pure CeCoIn$_5$ by Willers \textsl{et al.}.\cite{Willers2010}  

Figure\,\ref{Fig_3}\,(a) summarizes the LD results by showing the size of the LD (LD-size) at the $M_5$ edge as function of substitution, green dots for Cd and orange dots for Sn. The definition of the LD-size is given in the inset of Fig.\,\ref{Fig_3}\,(a). The advantage of the LD-size approach is that it is an observable and independent of any simulation. The data points show clearly two distinct trends: no change for all Cd doped samples and a decreasing and finally inverted LD-size for Sn substitutions. The dashed lines in Fig.\,\ref{Fig_3}\,(a) are guides to the eye. 

In order to obtain $\alpha^2$ from the LD-size, a full multiplet simulation is required (see Fig.\,\ref{Fig_4} and inset of Fig\,\ref{Fig_3}\,(a)) or the LD can be fitted with incoherent sums of the LDs of the pure $|\pm5/2\rangle$ and $|\mp\,3/2 \rangle$ states.\cite{Hansmann2008} The latter implies that the relationship between LD-size and $\alpha^2$ is linear. The value $\alpha^2$\,=\,1/6 corresponds to the cubic situation where the LD-size is zero; for $\alpha^2$\,$\textless$\,1/6 the $\Gamma_7$ orbital is elongated (prolate) along the $c$ direction because of the yo-yo shaped $|\pm3/2\rangle$ and donut-shaped $|\pm5/2\rangle$ charge densities (see bottom of Fig.\,\ref{Fig_3}\,(b)). Conversely, the $\Gamma_7$ charge density is compressed (oblate) for $\alpha^2$\,$\textgreater$\,1/6. Changes in the wave functions are small, in agreement with the small changes in the anisotropy of the static susceptibility.\cite{Bauer2006}
 
Nevertheless, results summarized in Fig.\,\ref{Fig_3} are unambiguous and imply that for all Cd substituted samples the 4$f$ orbital remains elongated in $c$ direction but it is shortened and finally compressed into the Ce-In(1) plane for an increasing level of Sn substitution. 

For comparison, Fig.\,\ref{Fig_3}\,(b) also includes $\alpha^2$ values for CeRhIn$_5$, CeIrIn$_5$, and the substitution series CeRh$_{1-\delta}$Ir$_{\delta}$In$_5$ that were investigated previously.\cite{Willers2015} In that study the LD-size and changes in it were larger so that correlations with the phase diagram could be detected more easily. In the CeRh$_{1-\delta}$Ir$_\delta$In$_5$ series, superconducting samples exhibit more elongated orbitals (i.e. smaller values of $\alpha^2$) compared to antiferromagnetic members.\cite{Willers2015} Independently, Fermi surface measurements have shown 4$f$ electrons contribute to a large Fermi surface in superconducting but not in antiferromagnetic samples,\cite{Haga2001,Fujimori2003,Harrison2004,Shishido2005,Settai2007,Shishido2007,Goh2008,Allen2013} i.e. $cf$ hybridization is stronger in the former ones (see purple arrow Fig.\,\ref{Fig_3}\,(b)). In view of anisotropy in 4$f$ orbitals found by XAS, the large Fermi surface of superconducting samples is a consequence of direction-dependent $cf$ hybridization, in agreement with theoretical predictions that $cf$ hybridization is strongest between the 4$f$ and out-of-plane In(2) wave functions.\cite{Shim2007,Burch2007,Haule2010} That is, the more elongated the $\Gamma_7$ orbital, the stronger the hybridization and superconductivity is promoted. These findings suggest CeCoIn$_5$ is the most strongly hybridized member of the Ce$M$In$_5$ series, in line with the fact that it has the highest superconducting transition temperature in the 115 family (see Ref.\,\onlinecite{Willers2015}).

In the present study, Cd substitution changes the ground state from superconducting to antiferromagnetic but the 4$f$ wave function remains globally unaltered. Sn substitution suppresses superconductivity and promotes $cf$ hybridization\,\cite{Bauer2005a,Daniel2005,Bauer2005,Donath2006,Bauer2006,Ramos2010} and here we find the orbitals become increasingly compressed (see orange arrow in Fig.\,\ref{Fig_3}\,(b)). 

The different behavior of the Cd and Sn substitution series as well as the opposite direction of the two arrows in Fig.\,\ref{Fig_3}\,(b) seem puzzling at first but, in fact, are quite reasonable when considering 1) the different effects of Cd and Sn substitution and 2) the different direction dependencies of hybridization due to substitution on the In site as in CeCo(In$_{1-y}$Sn$_y$)$_5$ compared to replacement of Rh by Ir as in CeRh$_{1-\delta}$Ir$_\delta$In$_5$. 

The results of the NQR investigation of CeCo(In$_{1-x}$Cd$_x$)$_5$ and CeCo(In$_{1-y}$Sn$_y$)$_5$ by Sakai \textsl{et al.}\,\cite{Sakai2015} concluded that Cd dopants on the In(2) sites locally enhance the magnetization, thus nucleating localized spins on neighboring Ce sites which then, eventually, form long-range magnetic order with increased Cd content,  i.e., the impact of Cd on the electronic structure is local while the majority of electrons remain unchanged  with Cd substitutions. Consequently, we observe no change in the wave function with Cd substitutions. This is contrasted by Sn substitutions that enhance the $cf$ hybridization homogeneously and with long-range spatial coherence . 

The Sn dopants occupy preferentially the in-plane In(1) sites.\cite{Daniel2005} In this case in-plane (and not out-of-the plane) hybridization is enhanced and as a result the extension of the $f$-orbitals is expected to increase within the tetragonal plane. This corresponds to larger $\alpha^2$ values [see the orbitals for $\alpha^2$\,=\,0 and 1 at bottom of Fig.\ref{Fig_3}\,(b)] and is in agreement with our observation. However, in CeIrIn$_5$ the Ir ions hybridize preferentially with the out-of-plane In(2) sites, i.e. in CeRh$_{1-\delta}$Ir$_{\delta}$In$_5$ the orbitals extend increasingly out of the $ab$-plane with increasing Ir concentration, and taller orbitals refer to smaller $\alpha^2$ values.  Hence, the different direction dependence of hybridization in CeCo(In$_{1-y}$Sn$_y$)$_5$ and CeRh$_{1-\delta}$Ir$_\delta$In$_5$ explains the opposite directions of the two arrows in Fig.\,\ref{Fig_3}\,(b).

\section{Conclusion}
The present study not only strongly supports the interpretation of nuclear resonance in terms of inhomogeneous versus homogeneous effects on the electronic structure by doping with Cd or Sn, respectively. It also shows that the spatial configuration of the 4$f$ wave function gives clues about the $cf$ hybridization that can otherwise only be inferred from band structure calculations and/or ARPES measurements. It is generally understood that the extent of $cf$ hybridization controls the nature of the ground state but differences in hybridization frequently are too small to be apparent in occupations of the 4$f$ shell.\cite{Sundermann2016} As these XAS measurements show, however, knowledge of the 4$f$ wave functions contributes to a deeper understanding of the driving mechanism for ground state formation.   

\section{Acknowledgments}
K.C., M.S., and A.S. benefited from the financial support of the Deutsche Forschungsgemeinschaft under project SE\,1441. Work at Los Alamos was performed under the auspices of the U.S. DOE, Office of Basic Energy Sciences, Division of Materials Sciences and Engineering, Z.F. acknowledges support of grant NSF-DMR-1708199.

\end{document}